\documentclass[useAMS,usenatbib]{mn2e}
\usepackage{amssymb}
\usepackage{amsmath}
\usepackage{graphicx}
\usepackage[draft]{hyperref} 
\usepackage{color}
\usepackage{soul}
\usepackage[utf8]{inputenc}

\begin{document}

\title[Calibrating the cosmic distance scale ladder]{Calibrating the cosmic distance scale ladder: the role of the sound horizon scale and the local expansion rate as distance anchors}

\author[A. J. Cuesta et al.]{\parbox{\textwidth}{
   Antonio J.~Cuesta\thanks{Email: ajcuesta@icc.ub.edu}$^1$,
   Licia Verde$^{2,1,3}$,
   Adam Riess$^{4,5}$,
   Raul Jimenez$^{2,1,6}$
 }
\\
$^1$ICC, University of Barcelona, IEEC-UB, Mart{\'\i} i Franqu{\`e}s 1, E-08028, Barcelona, Spain \\
$^2$ICREA (Instituci\'o Catalana de Recerca i Estudis Avan\c{c}at) \\
$^3$Institute of Theoretical Astrophysics, University of Oslo, 0315 Oslo, Norway \\
$^4$Department of Physics and Astronomy, Johns Hopkins University, 3400 North Charles Street, Baltimore, MD 21218, USA \\
$^5$Space Telescope Science Institute, 3700 San Martin Drive, Baltimore, Maryland 21218, USA \\
$^6$Institute for Applied Computational Science, Harvard University, MA 02138, USA \\
}

\maketitle

\begin{abstract}
We exploit cosmological-model independent measurements of the expansion history of the Universe to provide a cosmic distance ladder. These are supernovae type Ia used as standard candles (at redshift between 0.01 and 1.3)  and baryon acoustic oscillations (at redshifts between 0.1 and 0.8) as standard rulers.
We calibrate (anchor) the ladder in two ways: first using the local $H_0$ value as an anchor at $z=0$ (effectively calibrating the standard candles) and secondly using the cosmic microwave background-inferred sound-horizon scale as an anchor (giving the standard ruler length) as an inverse distance ladder. Both methods are consistent, but the uncertainty in the expansion history $H(z)$ is smaller if the sound horizon scale is used. 
We present inferred values for the sound horizon at radiation drag $r_d$ which do not rely on assumptions about the early expansion history nor on cosmic microwave background measurements but on the cosmic distance ladder and baryon acoustic oscillations measurements. 
We also present derived values of $H_0$ from the inverse distance ladder and we show that they are in very good agreement with the extrapolated value in a $\Lambda$CDM model from \textit{Planck} cosmic microwave background data.
\end{abstract}


\section{Introduction}

Accurate distance determinations at cosmological distances have been one of the observational evidences on which the standard cosmological model is built. It is the distance redshift relation that gives us the Universe's expansion history and from there we gather information about the Universe's content (dark matter and most importantly dark energy) e.g., \cite{Riess98, Perlmutter99}.
 
Since no one technique could, until recently, measure distances of extra-galactic or cosmologically distant objects\footnote{But see \citet{Simon2005,Riessparallax,Masers1,Masers2} for recent advances on this front.}, a succession of methods was used. In this approach -- the cosmic distance ladder-- each rung of the ladder provides the information necessary to determine the distance of the next rung, see e.g., \citet{MRR} for an historical introduction.
Traditionally, the cosmic distance ladder relies on standard candles and in particular type 1a supernovae, to extend the ladder well into the Hubble flow,  i.e., at distances  beyond roughly 100 Mpc.
Type 1a supernovae (SN1a) are still today one of the key datasets to map the expansion history of the universe at $z\lesssim1$ \citep[e.g.,][]{Betoule2014, Riessetal07, Hickenetal09, Conleyetal11, Suzukietal12, Sakoetal14}. On their own however they only provide an ``uncalibrated" distance scale as the absolute magnitude of the standard candle  cannot be accurately modeled or derived from theory. In other words the (relative) distance scale they provide must be calibrated, and this is  traditionally done with a distance ladder.
 
Since 2005 another technique to measure extragalactic distances has become possible \citep{EisensteinBAO2005,ColeBAO2005} and it is called Baryon Acoustic Oscillations (BAO). In the past ten years BAO measurements have undergone a spectacular development having now been measured from several independent surveys with few percent precision over the redshift range from $z=0.1$ to $z>1$ \citep[see e.g.,][]{Anderson2014,Tojeiro2014, Font2014, Delubac2014, Beutler2011, Ross2014, Kazin2014}. This is a standard ruler technique: the length of sound horizon at recombination is imprinted in the clustering of dark matter and its tracers like galaxies, provided one can accurately model the possible evolution of the observational signature due to gravitational instability. The distance scale given by the BAO feature as measured from large-scale structure must also be calibrated by knowing the size of the standard ruler. This is provided by Cosmic Microwave Background (CMB) observations.  The sound horizon determination from CMB data is somewhat cosmological model-dependent, as it is sensitive to the early expansion history and the composition of the early universe. Nevertheless it is exquisitely well measured-- its error-bar being below 1 per cent --  for models with standard early expansion history (e.g., standard number of effective neutrino species etc.), and extremely robust to systematic and instrumental errors \citep[see e.g.,][]{Planck2013}. However a recent model-independent determination of the standard ruler size is given by \cite{Heavens2014}.

If standard candles, calibrated from the local measurement of $H_0$ provide a {\it ``direct"} cosmic distance ladder (from nearby out towards cosmological distances), 
the BAO provides an {\it ``inverse"} cosmic distance ladder, calibrated at recombination $z\sim 1100$ and extended in, towards lower redshifts.

The spectacular progress in surveying the Universe of the past decade means that SN1a and BAO measurements now overlap in redshift and the statistical errors in both distance measures as function of redshifts are reaching percent level. 

\begin{figure*}
\includegraphics[width=\textwidth]{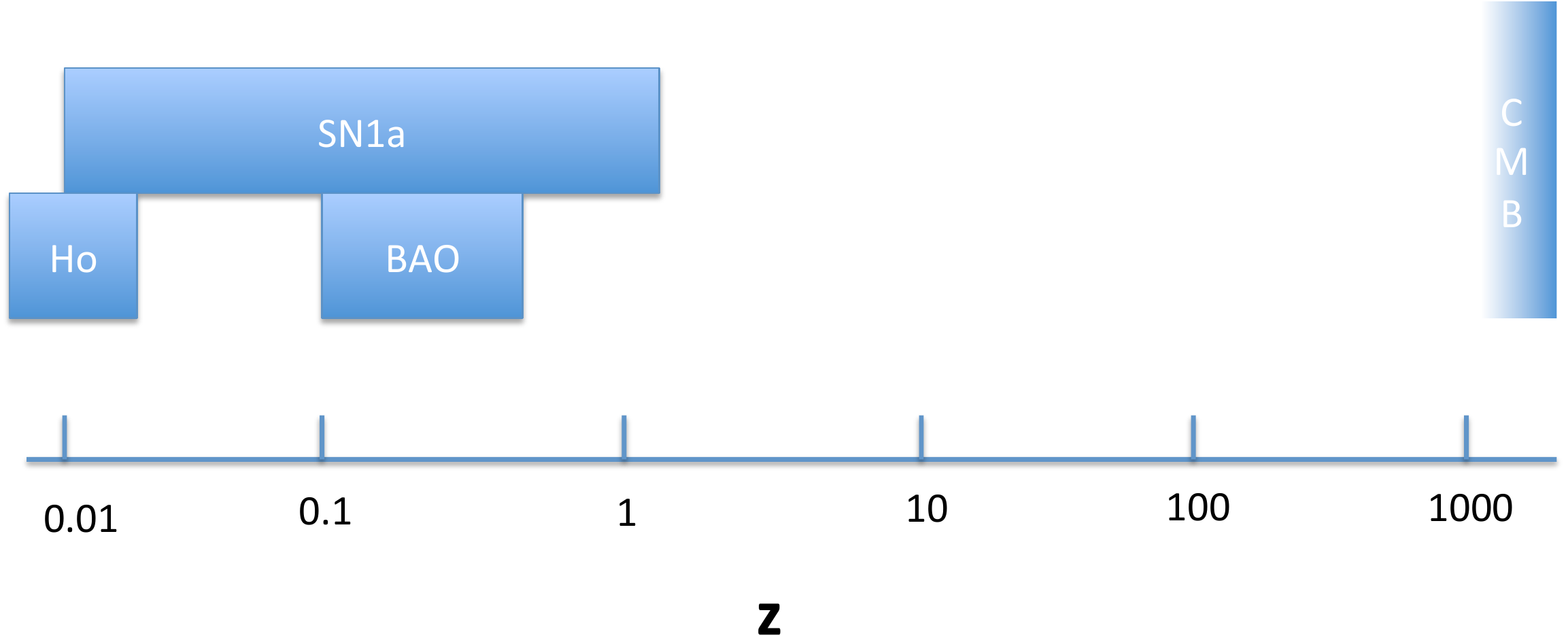}
\caption{ Schematic diagram of the redshift ranges covered by each of the datasets used in this paper. From low to high redshift we show the local measurement of the expansion rate $H_0$, the luminosity distances of type Ia supernovae, the distance determinations from the baryon acoustic oscillations in galaxy clustering, and the sound horizon scale in the cosmic microwave background.}
\label{fig:zranges}
\end{figure*}

This implies that now the {\it direct} and {\it inverse} cosmic distance ladders overlap and can be calibrated off one another  see Figure~\ref{fig:zranges}.

Here we consider the SN1a distance ladder and the BAO one first separately then jointly. This paper shows a similar approach to that shown in \cite{BOSSBAO} and \cite{Heavens2014}, in the sense that they also combine BAO and SN1a to build a distance ladder. However, there are differences. In \cite{BOSSBAO} only the \textit{inverse} cosmic distance ladder is considered to derive the value of the Hubble constant, so there is no attempt to infer the value of the sound horizon scale, which we do here. The approach in \cite{Heavens2014} is different in aims and underlying assumptions. While we work in the framework of the $\Lambda$CDM model and its extension, their analysis only relies on the assumptions of homogeneity and isotropy, a metric theory of gravity, a smooth expansion history, and the existence of standard candles (SN1a) and a standard BAO ruler (i.e., no dark energy modeling or General Relativity is assumed, only a Friedmann-Robertson-Walker metric). With only these assumptions, and using standard clocks as an additional dataset, they measure both the Hubble constant and the size of the standard ruler. They also explore the role that a prior on the Hubble constant (similar to the one we use here in the \textit{direct} distance ladder) plays on the results.
 
We begin by reviewing the basic equations and the  state of the art in Sec~2. Then in Sec.~3 we present the data sets  we use. In Sec.~4 we proceed to  first calibrate the cosmic distance ladder represented by the SN1a  to the sound horizon measurement provided by the CMB via the BAO measurements. This provides an inverse distance ladder and a derived determination of the Hubble constant. Then  we calibrate the SN1a +BAO distance ladder with the local $H_0$ measurement and infer  values for the sound horizon at radiation drag which are independent on early universe physics. Finally we  report constraints on the expansion history both absolute $H(z)$ and relative $E(z)=H(z)/H_0$. We draw our conclusions in Sec.~5.

\section{State-of the art and background}

From the above discussion it should be clear how important an absolute distance scale is and how this is directly related to the $H_0$ determination.
 In the era of precision cosmology, discrepancies of about 10 per cent, in supposedly well-know cosmological parameters such as the Hubble constant, have generated contradictory claims about being an indication for new physics that might explain the difference \citep[e.g.,][]{Bennett2014, Wyman2014, Verde2013, Marra2013, Verde2014}. Such is the case of the difference between direct measurements of the local expansion rate from \cite{Riess2011}  even after the recalibration of the distance to NGC 4258 \footnote{Hereafter  we refer to this recalibrated value simply as "Riess".}  from \cite{Humphreys2013}, 73.0$\pm$2.4 km s$^{-1}$ Mpc$^{-1}$, and the value of $H_0$ extrapolated assuming a $\Lambda$CDM model from the epoch of recombination to redshift $z=0$ from Cosmic Microwave Background (CMB) measurements by the {\it Planck} satellite \citep{Planck2013}, 67.3$\pm$1.2 km s$^{-1}$ Mpc$^{-1}$. It is important to stress that the CMB estimates of $H_0$ are extrapolations, and therefore are cosmological-model dependent.

\begin{figure*}
\includegraphics[width=0.49\textwidth]{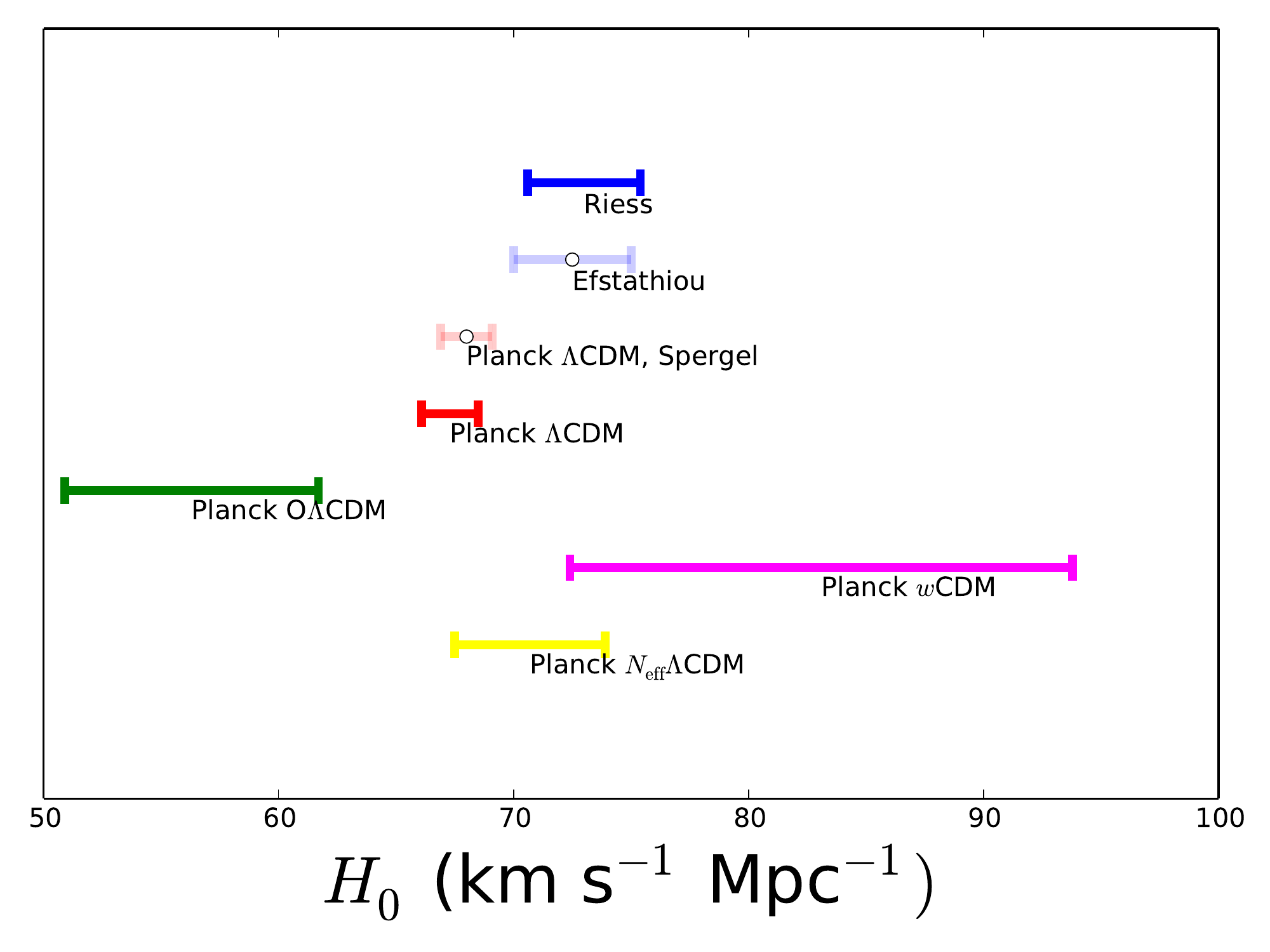}
\includegraphics[width=0.49\textwidth]{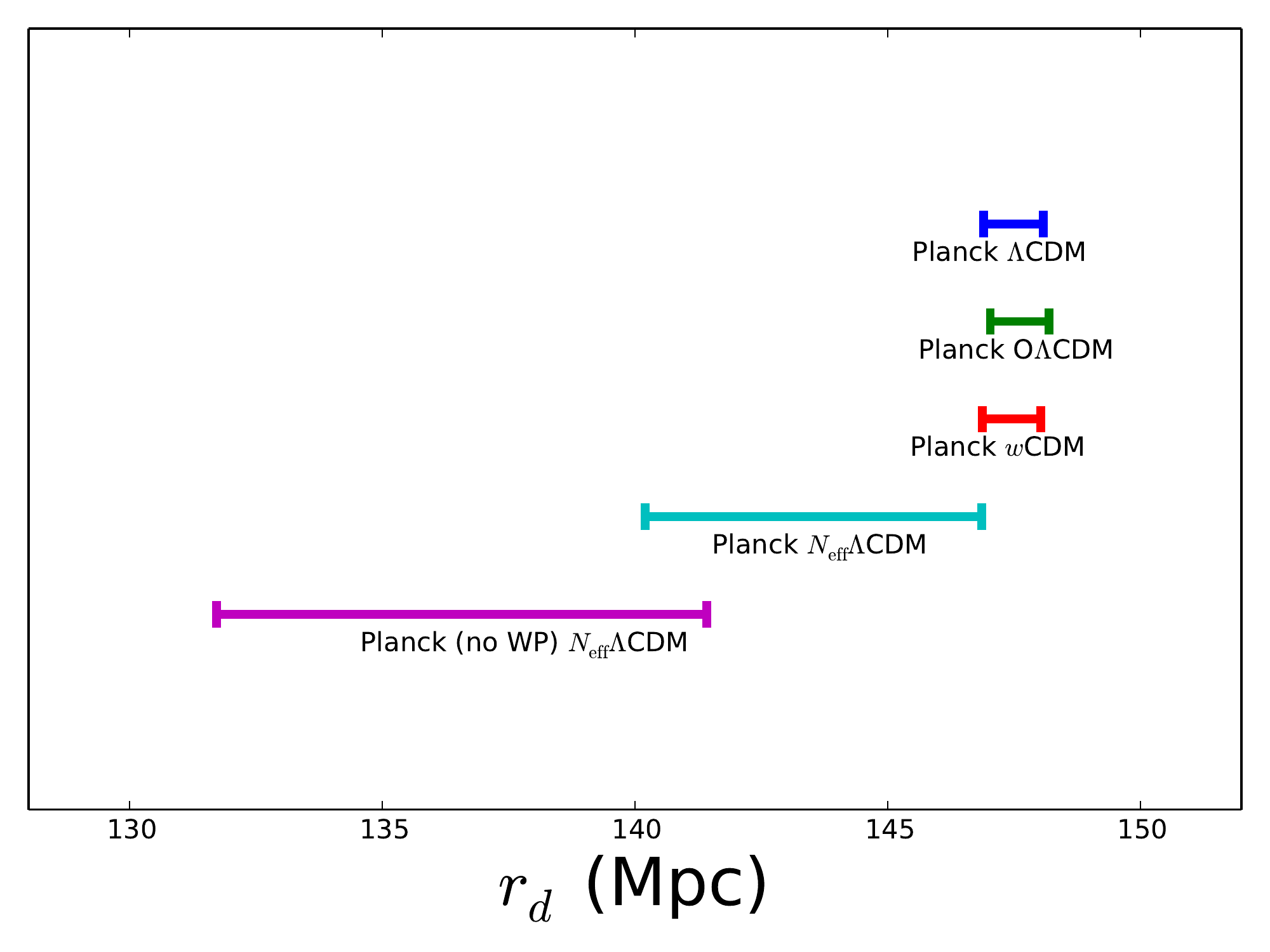}
\caption{Left panel: Comparison between local  measurements (using \citet{Riess2011,Humphreys2013} (dark blue, upper bar) and its reinterpretation by \citet{Efstathiou2014} (light blue, lower bar)) and CMB-derived measurements of the Hubble constant from {\it Planck+WP} data (labelled as Planck) for several assumed cosmologies. The error-bars correspond to 68 per cent confidence. The tension between the local and the CMB determinations is evident for some models ($\Lambda$CDM and O$\Lambda$CDM ) but not for others ($w$CDM or $ N_{\rm eff}\Lambda$CDM). The two measurements labelled ``Planck $\Lambda$CDM" refer to the Planck collaboration measurement (dark red, lower bar,\citet{Planck2013}) and the re-analysis of \citet{Spergel2013} (light red, upper bar).   The bars in faded out colors represent reinterpretations of the original datasets represented in solid colors. Right panel: the sound horizon scale: its determination is virtually cosmology-independent for cosmologies that differ on late-time history of the universe, but the determination is extremely sensitive to uncertainties in the early (pre-recombination) history.  }
\label{fig:barplot}
\end{figure*}

The state-of-the art in CMB data is provided by the temperature anisotropy measurements by the {\it Planck} satellite \citep{Planckoverview, Planck2013} which is almost always combined with the polarisation data at low multipoles from the WMAP satellite \citep{WMAP9} referred to as {\it WP}. The {\it Planck} data are often further complemented by higher multipoles measurements by the ACT and SPT experiments \citep{ACT, SPT} referred to as {\it highL}.
  
We can quantify how model dependent is the CMB-based $H_0$ determination 
using the publicly-released Monte Carlo markov chains (MCMC) ran by the {\it Planck} collaboration to explore the cosmological parameter space and find parameter estimates. We find that the {\it Planck+WP} dataset constraints $H_0$ to be within $64.2<H_0<70.4$ km s$^{-1}$ Mpc$^{-1}$ (at 99 per cent confidence level) for the flat $\Lambda$CDM model, whereas this becomes $43.3<H_0<71.4$ km s$^{-1}$ Mpc$^{-1}$ for the non-flat $\Lambda$CDM (O$\Lambda$CDM) model ($43.2<H_0<69.3$ km s$^{-1}$ Mpc$^{-1}$ when using {\it Planck+WP+highL}), and $58.7<H_0<100$ km s$^{-1}$ Mpc$^{-1}$ (with the upper boundary being set by the prior) for a model where the dark energy is not a cosmological constant but its equation of state parameter does not change in time (the $w$CDM model). Therefore a ``concordance" value of 70 km s$^{-1}$ Mpc$^{-1}$ \citep{Bennett2014} or the central measured value by \citet{Riess2011,Humphreys2013} 73.0 km s$^{-1}$ Mpc$^{-1}$ can be considered ruled out or perfectly acceptable depending on the context of the cosmological model (as already discussed in the literature see e.g., \cite{Verde2014}). This is summarised in the left panel of Fig.~\ref{fig:barplot} where 68 per cent confidence regions are shown.

 Re-analysis by \cite{Spergel2013} dropping the 217GHz data of the {\it Planck} dataset and  re-analysis of the direct distance ladder by \cite{Efstathiou2014}  report a modest shift  (less than 0.5 $\sigma$) in their Hubble constant determinations. This can be appreciated in Fig.~\ref{fig:barplot} left panel.

The CMB on the other hand, offers directly the absolute distance calibrator for the BAO, the sound horizon at radiation drag\footnote{This is slightly different from the sound horizon at recombination but the two quantities are tied to one another.}, $r_d$. This quantity is exquisitely well measured, yet it shows some small cosmological dependence. For example while it is measured with a 0.4 per cent uncertainty in the $\Lambda$CDM model for the {\it Planck+WP} data set ($r_d=147.49\pm0.59$ Mpc, \cite{Planck2013}), its central value is about 2.7 per cent, or 4 Mpc, lower in a model where the effective number of neutrino species, $N_{\rm eff}$, is not fixed to the standard value corresponding to three neutrino families, but is allowed to vary\footnote{Recall that  $N_{\rm eff}$ parameterises non-standard early expansion history}, $N_{\rm eff}\Lambda$CDM, $r_d=143.5 \pm 3.3$ Mpc \citep{Planck2013}. This is illustrated in the right panel of Fig.~\ref{fig:barplot}, where the error-bars correspond to 68 per cent confidence.

In summary, there is a residual cosmological dependence in $r_d$, which is however very mild when the late-time expansion history or the geometry is concerned. In these cases the standard ruler is measured with better than  per cent precision.
However, as expected, when the early expansion history is affected (as in the case with a possible dark-radiation component, illustrated by the $N_{\rm eff}$ case) the $r_d$ determination is degraded to a 2.3 per cent measurement (68 per cent confidence). It is important to bear in mind that a 3 per cent knowledge of the $r_d$ calibrator at $z\sim 1100$ is comparable to that of the $H_0$ one at $z=0$. In other words the inverse distance ladder calibration is significantly better than the direct one only  if the early expansion history is virtually fixed.

The Hubble parameter as a function of redshift $H(z)$ is the key quantity we seek to measure
\begin{equation}
H(z)=H_0 E(z)
\end{equation}
where, for example, for a non-flat Universe with generic equation of state parameter $w(z)$:
\begin{eqnarray}
E(z)&=&\left\{ \Omega_m (1+z)^3+\Omega_k(1+z)^2+\right. \nonumber \\ 
&&\left. \Omega_{\Lambda} \exp\left[3\int_0^z\frac{1+w(z')}{(1+z')}dz'\right]\right\}^{1/2}\,.
\end{eqnarray}

Here $\Omega_{\Lambda}$ and $\Omega_m $ denote the present day dark energy and dark matter densities normalised to the critical density; the curvature parameter is $\Omega_k=1-\Omega_m-\Omega_{\Lambda}$. 
Of course for a flat $\Lambda$CDM model we have:
\begin{equation}
E(z)= \sqrt{\Omega_m(1+z)^3+(1-\Omega_m)}\,.
\end{equation}
 
In practice SN1a measure the luminosity distance; each (unnormalised) standard candle at redshift $z$ can  ultimately yield an estimate of:

\begin{eqnarray}
d_L(z)&=&H_0D_L(z)=(1+z)H_0D_M(z)\\
\nonumber
&=&\frac{(1+z)}{\sqrt{\Omega_k}}{\rm sink}(\sqrt{\Omega_k}D(z))
\end{eqnarray}
where sink$(x)= \sinh(x),\, x$ or $\sin(x)$ if the curvature is negative, zero or positive respectively, $\Omega_k$ is the curvature parameter (in units of the critical density) and 
\begin{equation}
D(z)=\int_0^z\frac{1}{E(z')}dz'=H_0D_C(z)\,.
\end{equation}
For flat spatial geometry $ d_L(z)=(1+z)D(z)$. Clearly $H_0$ gives the normalisation.
In practice SN1a data constrain the distance modulus $\mu=m-M$,  the difference between the apparent and absolute magnitude of each SN1a:
\begin{eqnarray}
    \mu(z)&=& 25+5\log_{10}\left(D_L(z)\right)\nonumber \\
   &=& 25+5\log_{10}d_L(z)-5\log_{10}H_0
    \label{eq:mu}
\end{eqnarray}
where $D_L(z)$ is in Mpc. Since $H_0$ is not known a priori and the absolute magnitude of the standard candles $M$  cannot be accurately modeled or derived from theory, $\mu(z)$ is not a direct measurement of $H(z)$, however we note that the fine slicing of the redshift range (here we use all 31 bins of \cite{Betoule2014}) allows us to compute several relative distances $\mu(z_i)-\mu(z_j)=5\log_{10}(d_L(z_i)/d_L(z_j))$ which from the above equation are {\it independent} from $H_0$. So the shape of $E(z)$ is constrained whereas its overall normalization is not. 

\begin{figure}
\includegraphics[width=0.5\textwidth]{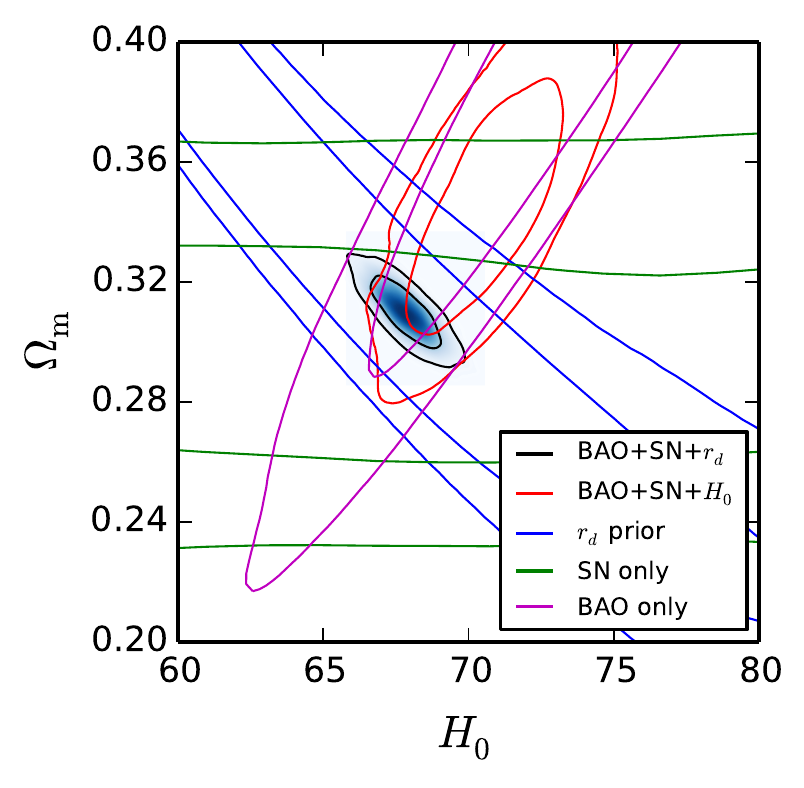}
\caption{Constraints in the $\Omega_m$--$H_0$ plane from BAO only, SN1a only, and the combinations BAO+SN1a+$H_0$ and BAO+SN1a+$r_d$. A $\Lambda$CDM model is assumed. Here $r_d$ is considered a derived parameter which depends on the densities of matter, baryons, and radiation, but we will drop that assumption in our analysis. The contours represent the 1 $\sigma$ and 2 $\sigma$ regions.}
\label{fig:contours}
\end{figure}

Most BAO analyses instead measure a combination of radial and angular signal $D_V/r_d$. 
\begin{eqnarray}
D_V(z)&=&\left[(1+z)^2D_A(z)^2 \frac{z}{H(z)}\right]^{1/3}=\left[ D_M(z)^2 \frac{z}{H(z)}\right]^{1/3}\nonumber \\
&=& \frac{1}{H_0}\left(z\frac{D(z)^2}{E(z)}\right)^{1/3}
\end{eqnarray}
and the sound horizon $r_d$ is (approximating for a matter dominated Universe at high redshift),
\begin{equation}
r_d=\frac{1}{H_0}\int_{z_d}^{\infty}\frac{c_s(z)}{E(z)}dz
\end{equation}

where $c_s(z)$ denotes the sound speed in the photon baryon fluid, $c_s(z)\simeq c/\sqrt{3(1+3\rho_b(z)/4\rho_r(z))}$ and $z_d$ the radiation drag redshift. Note that we have highlighted explicitly the $H_0$ dependence of $r_d$, however $z_d$ can be parameterised as a function of $\Omega_b h^2$ and $\Omega_m h^2$ \citep{EisensteinHu1998}, which together with the $\Omega_m$ dependence of $E(z)$   break the degeneracy and constrain $h$, only from BAO and SN1a data  if the baryon to radiation ratio is fixed. The baryon to photon ratio is exquisitely well measured by the CMB for all models with standard early expansion history. Therefore the dependence on $H_0$ is not completely eliminated in $D_V/r_d$. This is illustrated in Figure~\ref{fig:contours}, which shows, for a $\Lambda$CDM case, the different degeneracy directions of SN1a data (green line), BAO data (magenta line), and from a Gaussian prior in $r_d$ 147.49$\pm$0.59 Mpc (blue line). The latter shows the dependence of $r_d$ on $\Omega_m h^2$. We also compare in this figure the contraints from the combination of BAO+SN1a and a $H_0$ prior of 73.0$\pm$2.4km s$^{-1}$ Mpc$^{-1}$ (red line), as opposed to when BAO+SN1a are calibrated using the Gaussian prior in $r_d$ (black line and filled blue contours). In what follows we will indicate results obtained under this assumption (i.e., that $r_d$ is a derived parameter which depends on the densities of matter, baryons, and radiation) with the '*' symbol.  Conversely, one can infer $r_d$ ignoring its dependence on the matter, baryon, and radiation densities (i.e., as if it were an independent cosmological parameter) from BAO (with or without the addition of  SN1a) and without any input from the CMB, if the ladder is calibrated on a local measurement of $H_0$ and a parameterised form of the expansion history is used.

The uncalibrated standard ruler yields 
\begin{equation}
d_V(z)=D_V(z)/r_d= \left(z\frac{D(z)^2}{E(z)}\right)^{1/3}\widehat{r_d}^{-1}\,.
\end{equation}
where 
\begin{equation}
\widehat{r_d}=\int_{z_d}^{\infty}\frac{c_s(z)}{E(z)}dz=H_0r_d\,.
\end{equation}

From these equations it is clear that uncalibrated standard candles and rulers can only yield relative expansion history information i.e. $H(z)/H_0$. Moreover, to infer constraints on the expansion history $H(z)/H_0$ from $d_L(z)$ and $d_V(z)$ an underlying cosmological model must be assumed (for example, the curvature). Because of the integral nature of $D(z)$, while for a given $E(z)$ only an assumption about curvature is needed to relate $E(z)$ to $D(z)$, to invert the relation going from $D(z)$ to $E(z)$ requires assuming a functional form for $E(z)$. 
Rather than working with a model-independent form for $E(z)$ (like a polynomial, or some function specified by its values at certain redshift values), here we use a suite of cosmological models: $\Lambda$CDM, O$\Lambda$CDM, $w$CDM and $ N_{\rm eff}\Lambda$CDM. In this case the parameters describing the expansion history are the standard cosmological background parameters relative to that model.

In what follows sometimes we will have to assume a fiducial cosmology, we take the {\it Planck} best-fitting $\Lambda$CDM model, where $\Omega_m=0.315$, $\Omega_{\Lambda}=0.685$, $H_0=67.3$km s$^{-1}$ Mpc$^{-1}$, and $r_d=147.49$Mpc.

\begin{figure*}
\includegraphics[width=0.49\textwidth]{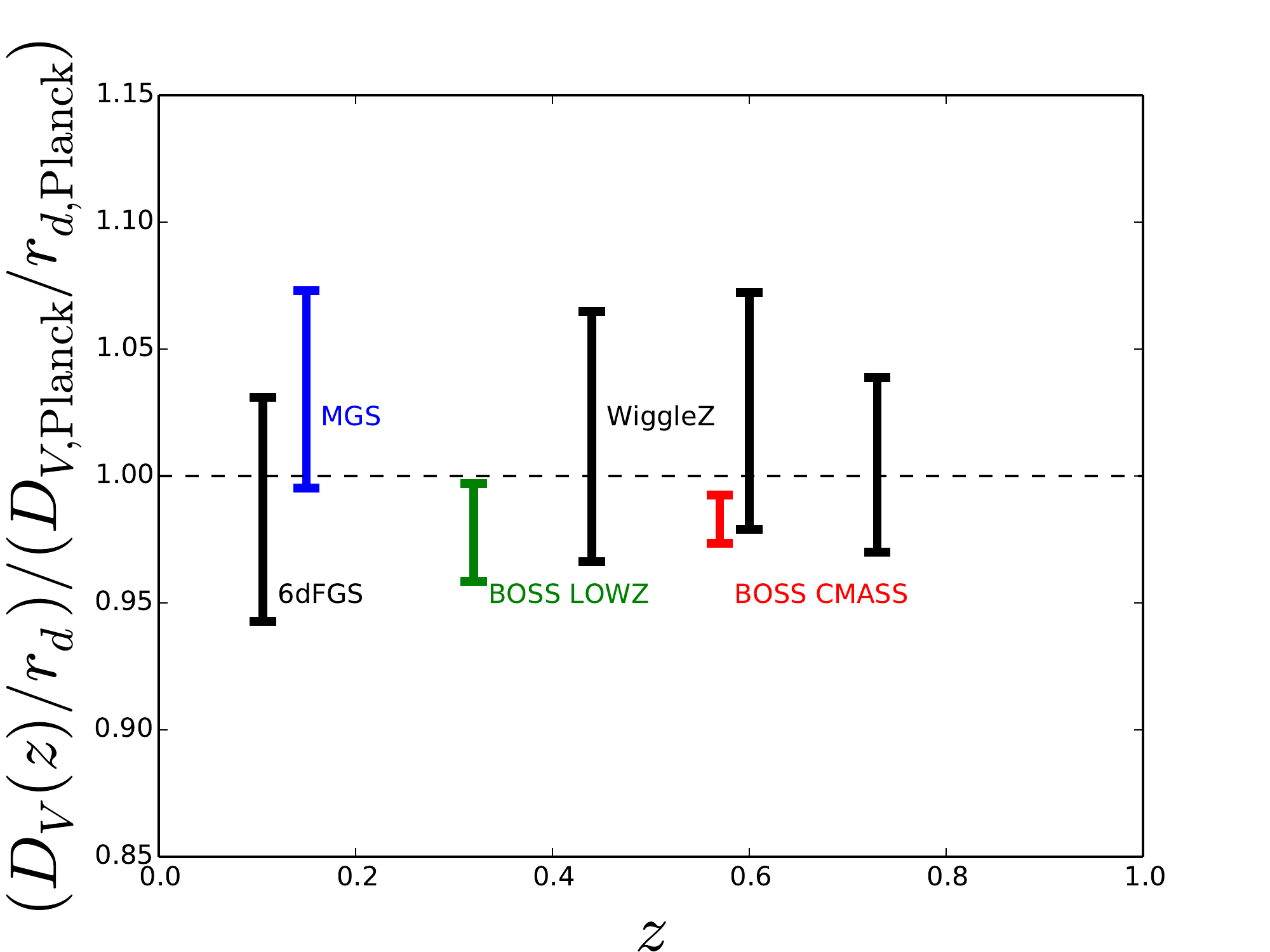}
\includegraphics[width=0.49\textwidth]{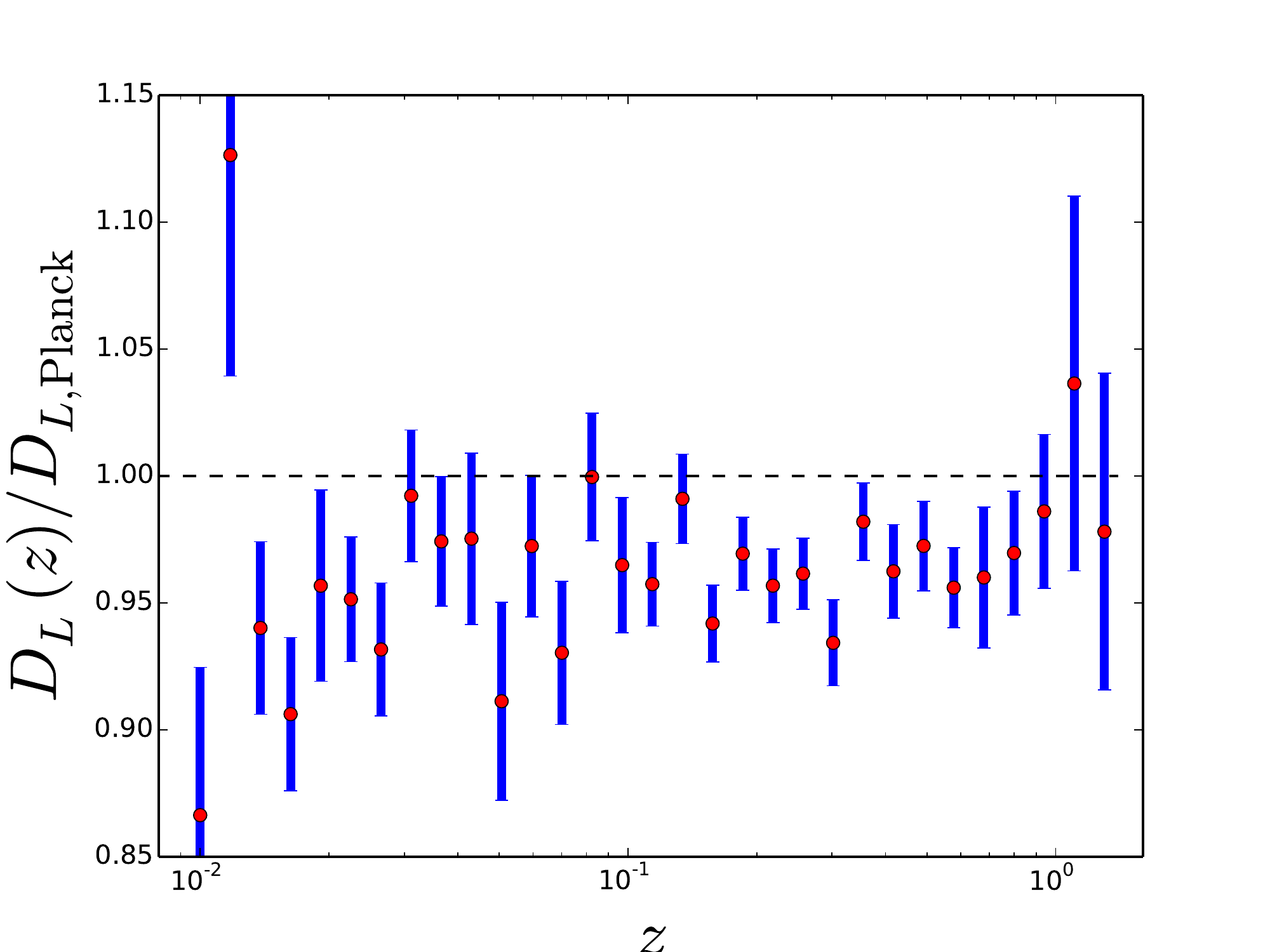}
\caption{Left: The distance-redshift relation as probed by current BAO measurements. The quantity plotted is $D_V(z)/r_d=((1+z)^2D_A(z)^2cz/H(z))^{1/3}/r_d$ normalised by the values for our fiducial cosmology given by the best-fitting parameters from the {\it Planck} analysis for a $\Lambda$CDM model BAO measurements shown in black are not used here, but are included in this plot for completeness. Right: the luminosity distance-redshift relation from SN1a measurements normalised by the fiducial cosmology values. Here the JLA sample has been binned using 31 nodes equally separated in log(1+$z$).  We remind the reader that these bins are correlated, therefore their full covariance matrix is included in our analysis and required to establish concordance with Planck.}
\label{fig:dz}
\end{figure*}

\section{Datasets}
\label{sec:data}
In this section we describe the cosmological datasets we use in this analysis. These are the recent BAO measurements from the SDSSIII BOSS survey \citep{BOSS} data release 11 and a recent compilation of Supernovae data which we describe in details below. A  compilation  of the state-of-the-art  galaxy BAO measurements is shown in the left panel of  Fig.~\ref{fig:dz} and the SN1a  measurements we use here are shown in  the right panel of Fig.~\ref{fig:dz}. Clearly most of the statistical power for the BAO (when used in conjunction with SN1a) comes from the two BOSS measurements, which we use here.

Both datasets are complementary in the sense that the distance measurements determined using BAO have high precision, but they  sparsely cover the redshift range. In particular, at low redshifts, due to the limited volume that can be observed, the error-bars are large. On the other hand, the SN1a compilation by \cite{Betoule2014} samples the redshift range $0.01<z<1.0$ really well. This gives a relative distance measurement (this is shown in Figure~\ref{fig:snvsbao}), with the normalization being unknown. Supernovae are usually normalised at $z=0$ using $H_0$ and BAO at $z=1100$ using $r_d$. But since the two ``ladders" overlap, they can be calibrated off each other.

\subsection{BAO data}

The galaxy  BAO measurements shown in Fig. \ref{fig:dz}  are the 6dF measurement at low redshift \citep{Beutler2011}, 
$r_d/D_V(0.106)=0.336\pm0.015$,
the Main Galaxy Sample (MGS) BAO from SDSS-II \citep{Ross2014},
$D_V(0.15)/r_d=4.47\pm0.16$ 
the two measurements from galaxy BAO from the baryon oscillation spectroscopic survey (SDSSIII-BOSS) \citep{Anderson2014,Tojeiro2014},
$D_V(0.32)/r_d=8.465\pm0.175$ and
$D_V(0.57)/r_d=13.77\pm0.13$ 
and the reconstructed WiggleZ measurements of \citet{Kazin2014}
$D_V(0.44)(r^{\rm fid}_d/r_d)=1716\pm83$~Mpc, $D_V(0.60)(r^{\rm fid}_d/r_d)=2221\pm101$~Mpc, and $D_V(0.73)(r^{\rm fid}_d/r_d)=2516\pm86$~Mpc.

The BOSS measurement at $z=0.57$ used here is the anisotropic measurement presented in \cite{Anderson2014}, which measures $D_A(z)$ and $H(z)$ rather than $D_V(z)$. For simplicity, we only include those measurements with smaller uncertainties at a given redshift. Those are the MGS BAO and the two galaxy BAO measurements shown in the left panel of Fig.~\ref{fig:dz} in blue, green, and red respectively. We have also tested the effect of adding the anisotropic BAO results from the Lyman-$\alpha$ forest of BOSS by \cite{Font2014, Delubac2014}. Since they do not change our results significantly, we will not include them here.

We have tested the consistency between the above  compilation of  BAO measurements and a $\Lambda$CDM model as described by the Planck best-fitting cosmological parameters. To do so we adopt the approach proposed in \cite{Verde2013,Verde2014} of measuring the multi-dimensional {\it Tension} ($T$) and interpret it in  terms of odds using the Jeffreys' scale.  In Table~\ref{tab:odds} we report $\ln T$ and  the odds that a set of BAO distance measurements (starting from low to high redshift) are consistent with Planck $\Lambda$CDM cosmology.  In the Jeffreys' scale odds less that 1:3 or $\ln T <1$ indicate that  there is  no indication of inconsistency. Strong or highly significant tension  would need odds $<1:12$ and $<1:150$.

\begin{table}
\caption{Odds (Cumulative in redshift up to $z_{\rm max}$) that BAO measurements are consistent with Planck $\Lambda$CDM cosmology.}
\begin{tabular}{l|c|c|}
\hline
$z_{\rm max}$ &  $\ln T$ &Odds\\
\hline
0.20   &    0.231496& 1:1.2\\
0.35   &    0.294514 &1:1.3\\
0.44   &    0.327827 &1:1.4\\
0.57   &    0.531189 &1:1.7\\
0.60   &    0.790740 &1:2.2\\
0.73   &   0.805645 &1:2.2\\
\hline
\end{tabular}
\label{tab:odds}
\end{table}

 For (isotropic) BAO, the distance measurements is encoded in terms of the angle-averaged distance $D_V(z)$. Being a combination of  $D_A(z)$ and $H(z)$  converting this type of measurement to a pure constraint on  $H(z)$ or a pure constraint on  $D_A(z)$  or  a direct comparison with the supernova measurements of  $D_L(z)$ (but see \cite{Lampeitl2010}), requires the assumption of a particular cosmological model, i.e., a shape of the expansion history. $H(z)$ is  particularly sensitive to changes in curvature and dark energy,  at the redshifts probed by BAO from galaxy clustering and SN1a  samples.

\subsection{SNe data}
The compilation of 740 Type Ia Supernovae by \cite{Betoule2014} comprises 239 supernovae by the SuperNovae Legacy Survey (SNLS) and 374 from the Sloan Digital Sky Survey (SDSS) as well as 118 supernovae from low-redshift surveys and a few (9) of them beyond $z>1$ observed using the HST. These are binned in 31 bins equally spaced in $\log(1+z)$ as in the appendix of  \cite{Betoule2014}.

The distance information from supernovae data is encoded in terms of the distance modulus $\mu(z)$, (see Eq.\ref{eq:mu}) 
implying that there is a one-to-one relation between $\mu(z)$ and the luminosity distance $D_L(z)=(1+z)^2D_A(z)$ for a given value of  $M$.

This relation has been calibrated in \cite{Betoule2014} (see Appendix E and Tables F1 and F2) from apparent magnitude of their SN1a compilation together with the color terms, shape of the light curve terms, and nuisance parameters, in an unbiased manner.  We marginalise over $M$ as in the  SN1a JLA module provided by \cite{Betoule2014}.

\section{Calibrating the Cosmic Ladder}
We consider the following models for the expansion history $\Lambda$CDM, O$\Lambda$CDM, $w$CDM and $ N_{\rm eff}\Lambda$CDM. Thus $E(z)$ is described by 1 (for $\Lambda$CDM) or 2 (other models) parameters and $H(z)$ depends on one extra parameter $H_0$.
We use the publicly available code {\sc CosmoMC} \citep{Lewis2002} to run Monte Carlo Markov chains and explore the posterior distributions and cosmological constraints for the SN1a and BAO datasets described in \S \ref{sec:data}. We include BAO from LOWZ and CMASS as implemented in the current version of the code and also the SN1a JLA module provided by \cite{Betoule2014} at the website \url{http://supernovae.in2p3.fr/sdss_snls_jla/ReadMe.html}. 
We explore a complete set of cosmology runs (see Table~\ref{tab:runs}), in which we combine BAO+SN1a, BAO+$r_d$ (BAO+ a CMB derived $r_d$ prior) and BAO+SN1a+$r_d$. In the case of $\Lambda$CDM we also explore the cosmological constraints from BAO and SN1a on their own.

\subsection{The Hubble constant and the inverse distance ladder}

\begin{figure*}
\includegraphics[width=0.49\textwidth]{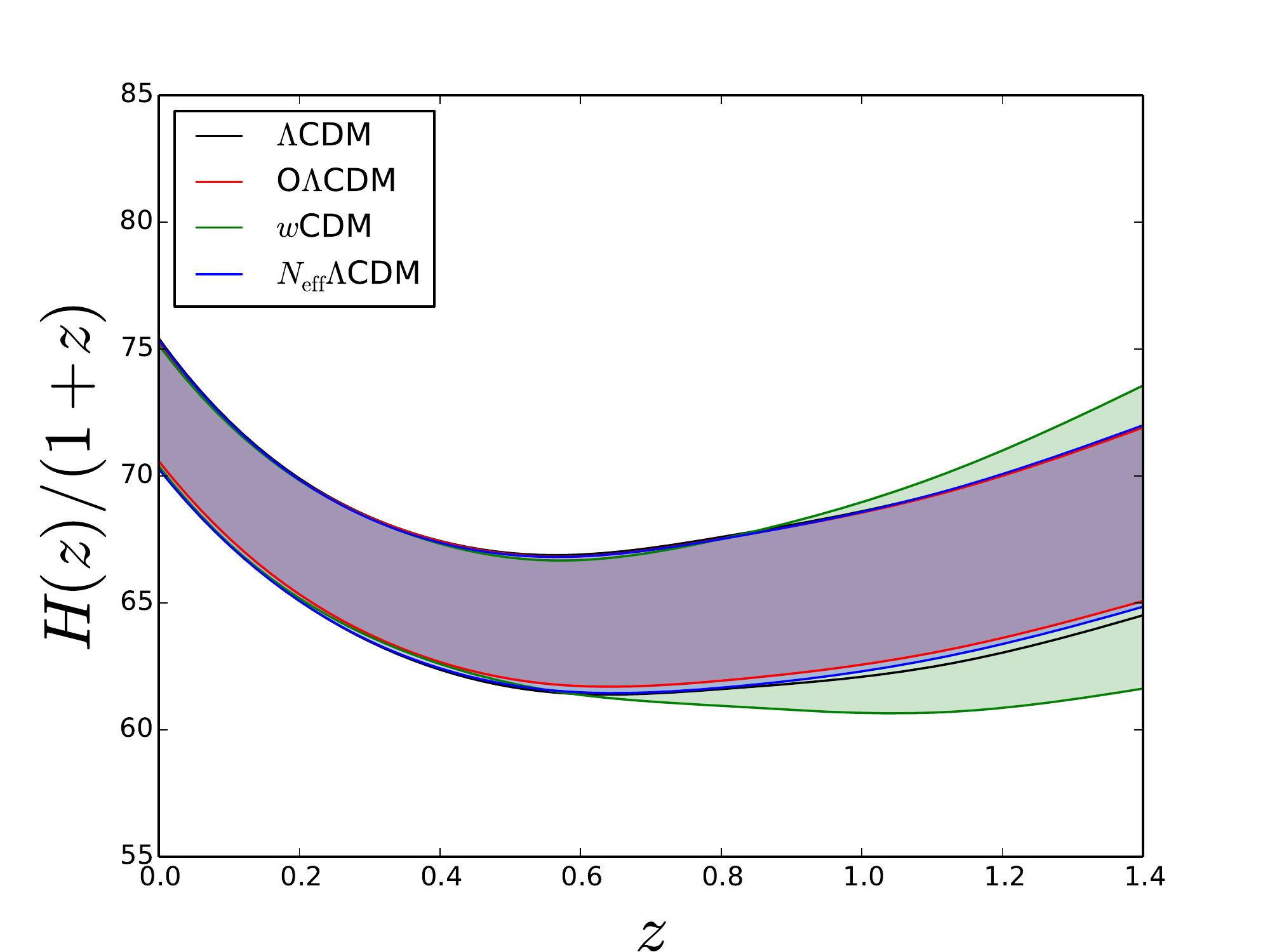}
\includegraphics[width=0.49\textwidth]{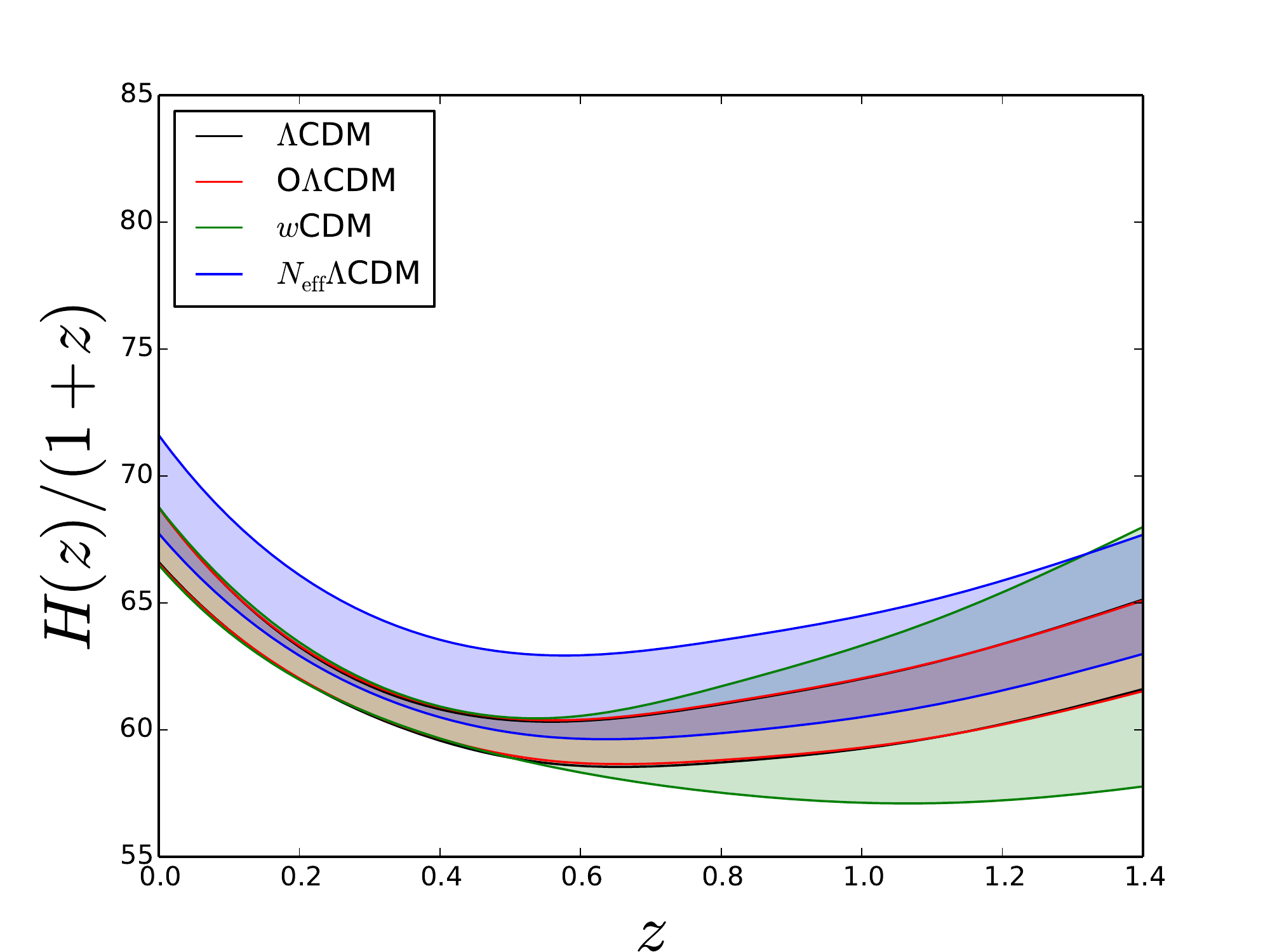}
\caption{Expansion history from  BAO+SN1a+$H_0$ (left panel) and BAO+SN1a+$r_d$ (right panel). Contours enclose the 68 per cent region of possible values of $H(z)$ at that $z$.}
\label{fig:prior}
\end{figure*}

To calibrate the BAO on the sound horizon scale for each of the models considered we use the corresponding Planck prior on $r_d$ (shown in the right panel of Figure~\ref{fig:barplot} and in the Planck+WP column of Tab. \ref{tab:rs}). Results are reported in Tab. \ref{tab:h0}. These results are also shown in the right panel of Figure~\ref{fig:prior}.  Note that  the determination in the $\Lambda$CDM model   rules out a Hubble constant of 74km s$^{-1}$ Mpc$^{-1}$, and is therefore somewhat in tension with the local determination of \cite{Riess2011,Humphreys2013}, as already pointed out before in the literature. Late-time changes to the expansion history ($w$CDM, O$\Lambda$CDM)  do not change this conclusion but early changes (see the $N_{\rm eff}$ case) do. This is at the core of the recent  proposals for a new concordance model with sterile neutrinos (e.g., \cite{Hamann2013, Wyman2014, Dvorkin2014, Battye2014}).

A more general analysis is found in Section 4 in \cite{BOSSBAO} by the BOSS collaboration in which they present $H_0$ constraints for more general cosmological models. We refer the reader to that paper for more details.

\begin{table*}
\caption{Comparison between Hubble constant values for different datasets and cosmological models. Units are km s$^{-1}$ Mpc$^{-1}$. The * denotes the case where $r_d$ is a derived parameter depending on the densities of matter, baryon and radiation. In this case, we fix the baryon and radiation densities to their best-fit Planck values.}
\begin{tabular}{l|ccc|c}
\hline
         $H_0$(km s$^{-1}$ Mpc$^{-1}$)      & BAO+SN* & BAO+$r_d$ & BAO+SN+$r_d$ & Planck+WP\\
\hline
$\Lambda$CDM  & 68.6$\pm$2.2 & 64.7$\pm$ 2.2 & 67.7$\pm$ 1.1 & $67.3 \pm 1.2$ \\
O$\Lambda$CDM & 64.5$\pm$7.4 & 64.8$\pm$ 2.2 & 67.6$\pm$ 1.1 & $56.3 \pm 5.4$ \\
$w$CDM & 72.5$\pm$11.1 & 66.1$\pm$ 2.3 & 67.7$\pm$ 1.1 & $83.1 \pm 10.7$\\
$ N_{\rm eff}\Lambda$CDM & 75.7$\pm$4.5 & 66.8$\pm$2.7 & 69.7$\pm$ 1.9 & $70.7 \pm 3.2$ \\ 
\hline
\end{tabular}
\label{tab:h0}
\end{table*}

\subsection{The sound horizon from the distance ladder and $H_0$}

In this Section rather than using a prior from the sound horizon $r_d$ we use the measurements from the local expansion rate $H_0$ to calibrate the standard ruler from the BAO on the (relative) distance vs. redshift relation from supernovae data.

\begin{figure}
\includegraphics[width=0.5\textwidth]{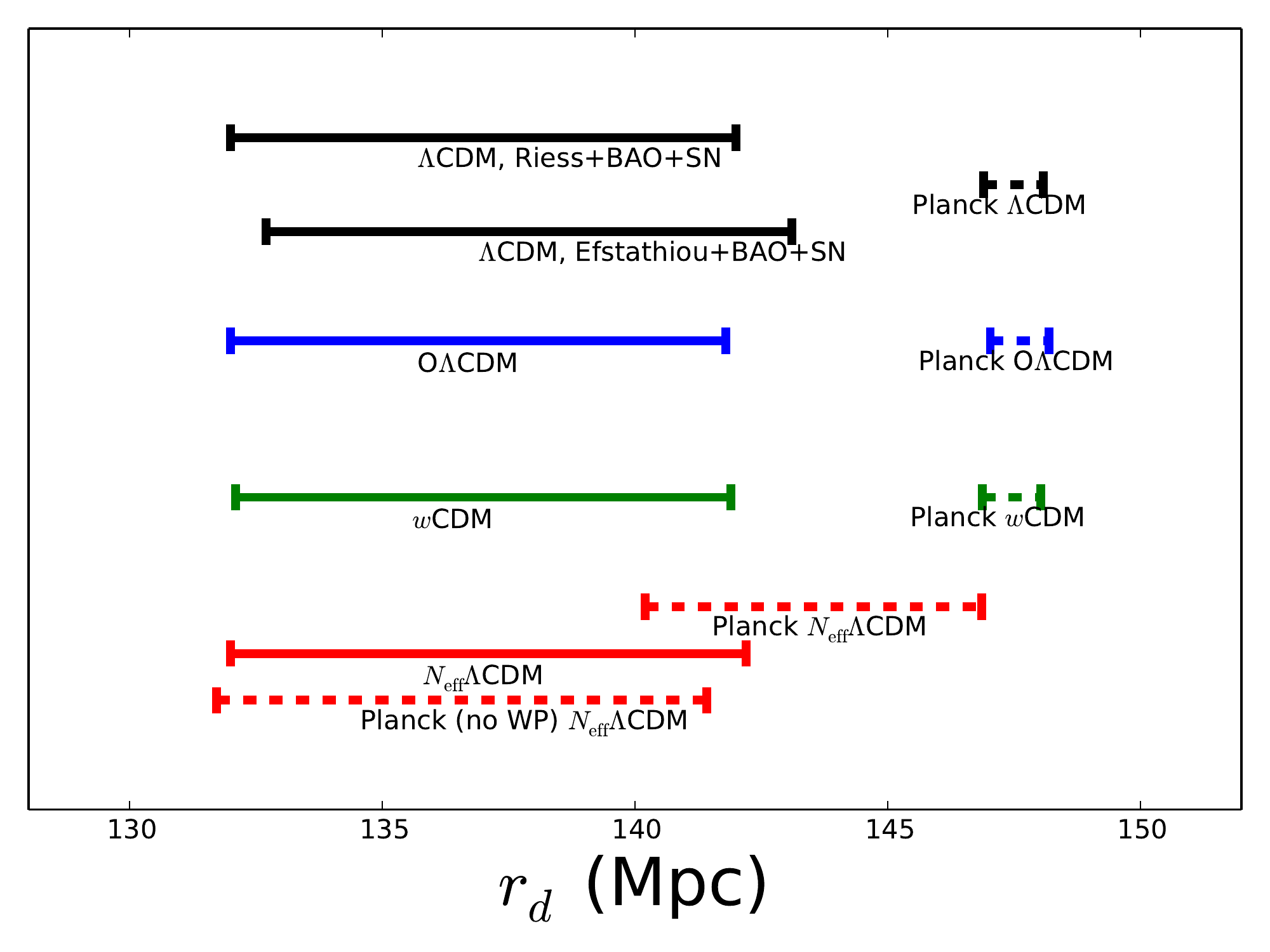}
\caption{Constraints on the sound horizon $r_d$ derived from SN1a+BAO+$H_0$ chains. Dashed lines show the constraints from CMB only, whereas solid lines show our results.}
\label{fig:rs}
\end{figure}

As shown in Figure~\ref{fig:rs}, once  we include a prior on the Hubble constant, the inferred distribution of the sound horizon scale is almost independent of the assumed cosmological model. 

We find values consistent with  Planck and WMAP9 measurements (see Table~\ref{tab:rs}). In this Table we also show the values, assuming a $\Lambda$CDM model, of the sound horizon scale when the Hubble constant prior is dropped (i.e. BAO+SN1a), but where we fix the radiation and baryon densities at their best-fitting values.   Note that the error on the  $r_d$ value  inferred from  the distance ladder  (and therefore insensitive to the early expansion history) is comparable to that  obtained from CMB measurements in the case of the $N_{\rm eff}\Lambda$CDM from Planck +WP data and smaller than that obtained from Planck data alone.

\begin{table}
\caption{Comparison assuming the $\Lambda$CDM cosmological model of the CMB measurement of the sound horizon $r_d$ and the direct measurement of the Hubble constant $H_0$ with extrapolations from BAO and SN data. The * denotes the case where $r_d$ is a derived parameter depending on the densities of matter, baryon and radiation. In this case, we fix the baryon and radiation densities to their best-fit Planck values.}
\begin{tabular}{l|cc}
\hline
Dataset & $r_d$ (Mpc) & $H_0$ (km s$^{-1}$ Mpc$^{-1}$) \\
\hline
Planck+WP & $147.49\pm 0.59$ & -- \\
Riess            & -- &$73.0\pm 2.4$ \\
\hline

BAO+SN+$H_0$ & 137.0$\pm$ 5.0 & 72.9$\pm$2.4 \\
BAO+SN+$r_d$ & 147.5$\pm$ 0.6 & 67.7$\pm$1.1 \\
BAO+SN* & 145.5$\pm$ 5.9 & 68.6$\pm$2.2 \\
BAO+$H_0$ & 132.1$\pm$ 5.7 & 72.6$\pm$ 2.4 \\
BAO+$r_d$ & 147.5$\pm$0.6 & 64.7$\pm$2.2 \\
SN+$H_0$ & $149\pm17$ & $73.0\pm 2.4$ \\
SN+$r_d$ & $147.5\pm0.6$ & $69.9\pm0.8$ \\
BAO* & 124$\pm$ 14 & 77.0$\pm$6.6 \\
SN* & $162\pm 27$ & unconstrained \\ 
\hline
\end{tabular}
\label{tab:lcdm}
\end{table}

\begin{table*}
\caption{Comparison between sound horizon values for different datasets and cosmological models. Units are Mpc. The * denotes the case where $r_d$ is a derived parameter depending on the densities of matter, baryon and radiation. In this case, we fix the baryon and radiation densities to their best-fit Planck values.}
\begin{tabular}{l|ccccc}
\hline
    $r_d$ (Mpc)  & BAO+SN*& BAO+$H_0$ & BAO+SN+$H_0$ & Planck+WP& Planck \\
\hline
$\Lambda$CDM & 145.5 $\pm$ 5.9 & 132.1$\pm$5.7 & 137.0$\pm$ 5.0 & $147.5\pm 0.6$ & $147.5\pm 0.6$  \\
O$\Lambda$CDM & 155.9 $\pm$ 16.2 & 132.0$\pm$5.8 & 136.9$\pm$ 4.9 &$147.6\pm 0.6$ & $147.6\pm 0.6$ \\
$w$CDM & 140.9 $\pm$ 22.0 & 132.5$\pm$6.3 & 137.0$\pm$ 4.9 &$147.5\pm 0.6$ & $147.5\pm 0.6$\\
$ N_{\rm eff}\Lambda$CDM & 132.0 $\pm$ 8.7 & 132.1$\pm$5.7 & 137.1$\pm$ 5.1 & $143.5\pm 3.3$  &$136.6\pm 4.9$ \\
\hline
\end{tabular}
\label{tab:rs}
\end{table*}

\begin{table*}
\caption{Cosmology runs studied in this paper.}
\begin{tabular}{l|ccccccc}
\hline
                          &  BAO only & SN only & BAO+SN & BAO+$H_0$ & BAO+$r_d$ & BAO+SN+$H_0$ & BAO+SN+$r_d$ \\
\hline
$\Lambda$CDM & Yes & Yes & Yes & Yes & Yes & Yes & Yes \\
O$\Lambda$CDM & No & No & Yes & Yes & Yes & Yes & Yes  \\
$w$CDM & No & No & Yes & Yes & Yes & Yes & Yes  \\
$ N_{\rm eff}\Lambda$CDM & No & No & Yes & Yes & Yes & Yes & Yes \\
\hline    
\end{tabular}
\label{tab:runs}
\end{table*}

\begin{figure*}
\includegraphics[width=0.49\textwidth]{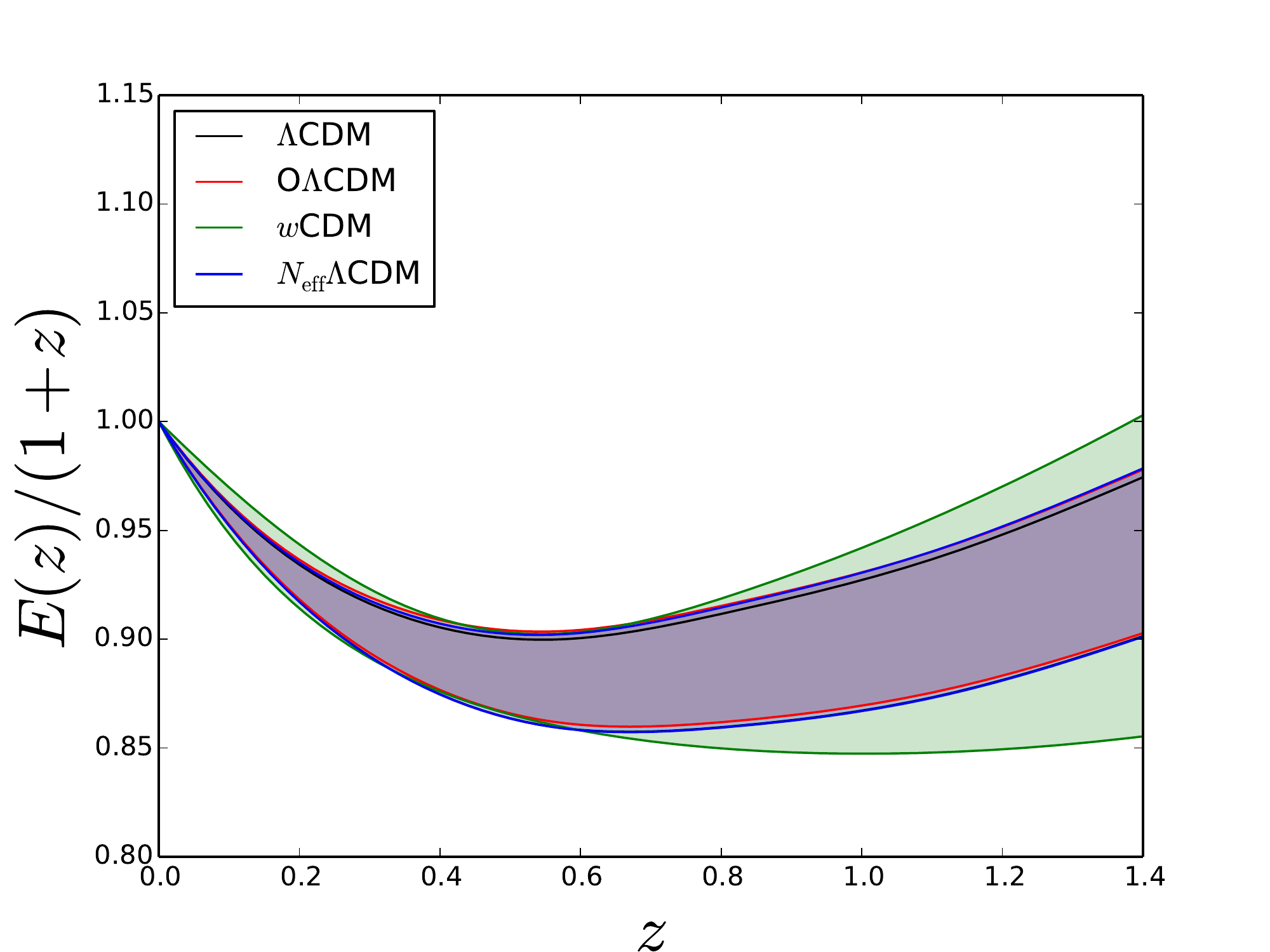}
\includegraphics[width=0.49\textwidth]{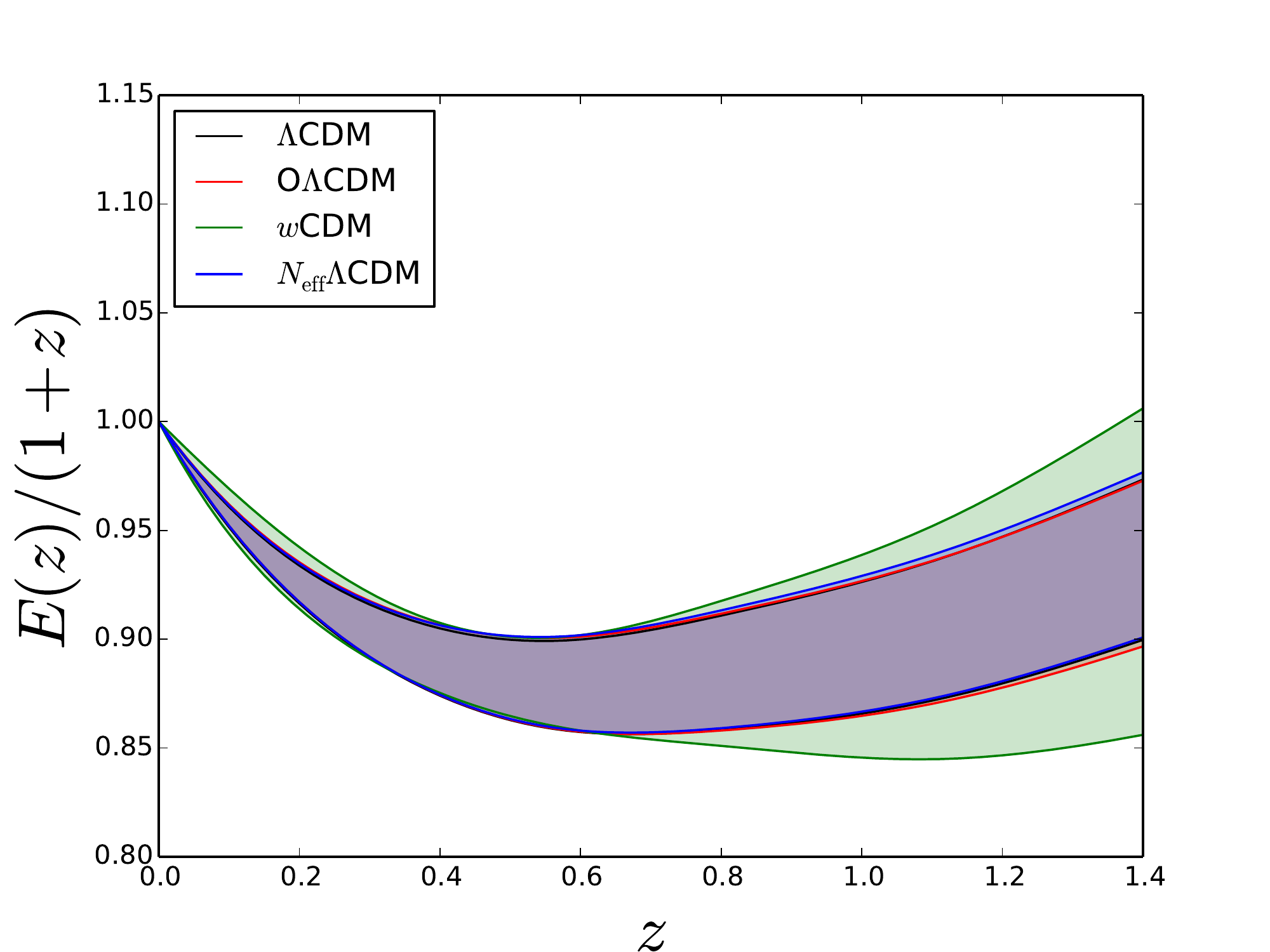}
\caption{Expansion history (normalised to $H_0$) from BAO+SN1a using the $\Lambda$CDM, $ N_{\rm eff}\Lambda$CDM, O$\Lambda$CDM and the $w$CDM cosmological models. Contours enclose the 68 per cent region of possible values of $H(z)/H_0$ at that $z$.}
\label{fig:ez}
\end{figure*}

\begin{figure*}
\includegraphics[width=0.49\textwidth]{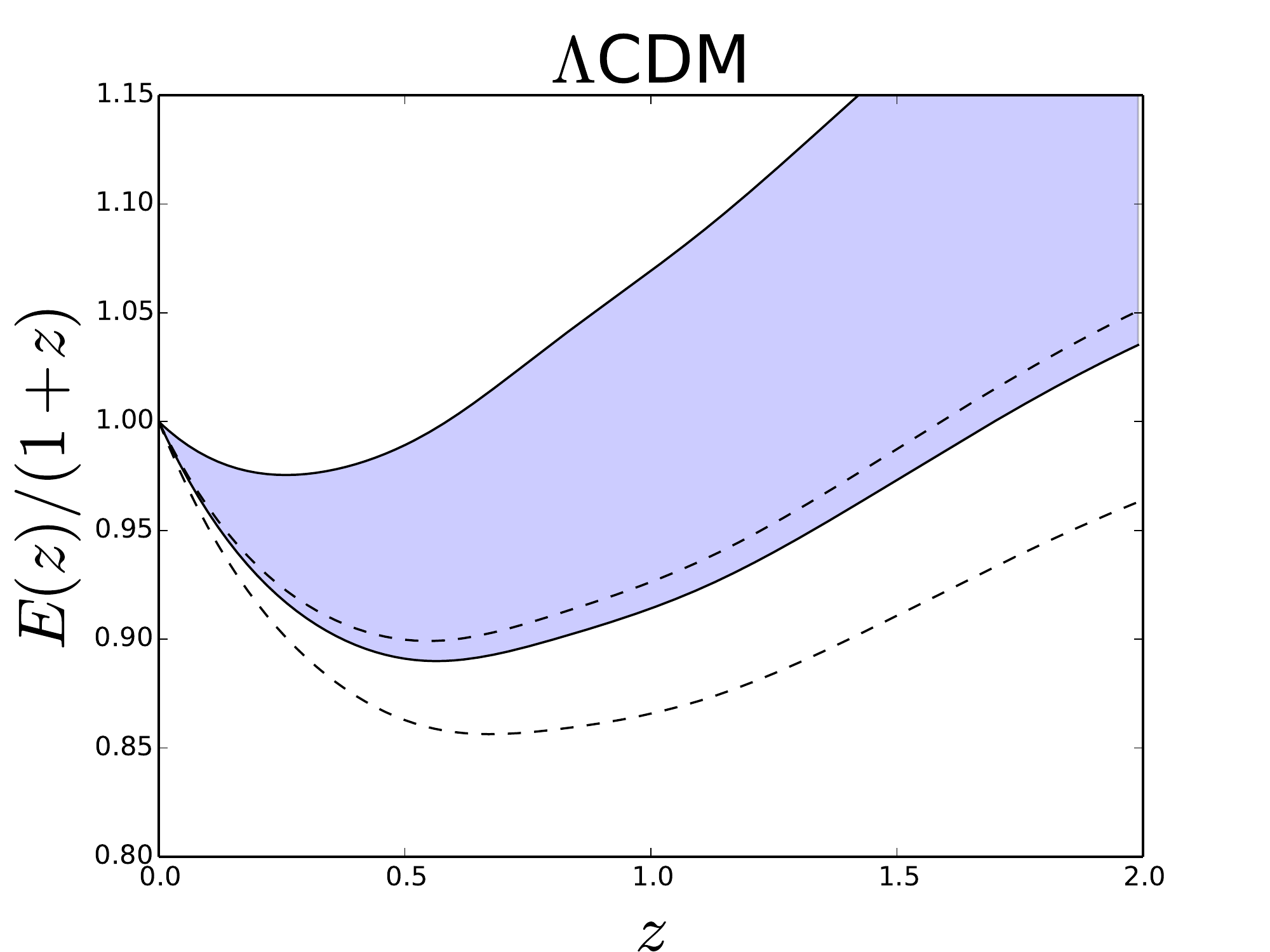}
\includegraphics[width=0.49\textwidth]{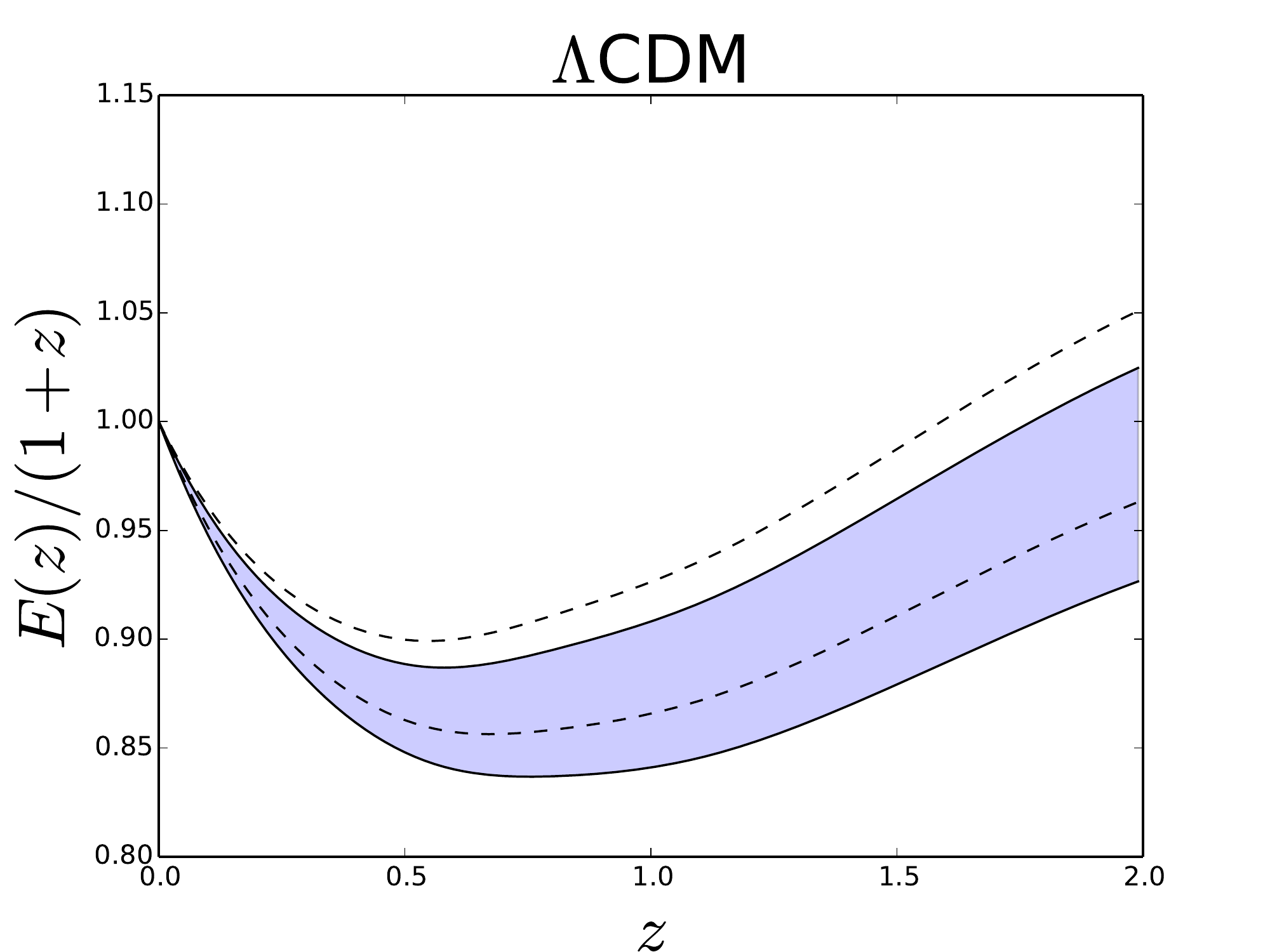}
\caption{Expansion history (normalised to $H_0$) from BAO only (left) and SN1a only (right) assuming a $\Lambda$CDM cosmological model. Contours enclose the 68 per cent region of possible values of $H(z)/H_0$ at that $z$. In both panels the dotted lines correspond to the BAO+SN1a+$r_d$ combination.}
\label{fig:snvsbao}
\end{figure*}

\subsection{Expansion history between $0<z\lesssim1$}

The expansion history of the Universe as derived by these intermediate redshifts cosmological probes is however much more dependent on the assumed cosmological model. In Figure~\ref{fig:prior} we show the derived expansion history for the $\Lambda$CDM cosmology. The quantity shown in the plot is the expansion rate $\dot{a}=H(z)/(1+z)$ as a function of redshift $z$.  Note how the uncertainty in the expansion history $H(z)$ depends on the error-bar on the distance calibrator $r_d$ (right panel) or $H_0$ (left panel).

On the other hand the shape of the expansion history i.e., $E(z)$ is much more robust to the underlying cosmology  as shown in  Fig.~\ref{fig:ez}. While with only two BAO measurements $E(z)$ is not well constrained even in the $\Lambda$CDM model (Fig. \ref{fig:snvsbao}),  the fine redshift sampling offered by SN1a yields a good determination of $E(z)$ over the full redshift range for the full set of models considered here.  In particular  the quantity $E(z)/(1+z)$ reported in Figure~\ref{fig:ez},   is useful to show the transition from a decelerating Universe when matter dominates to an accelerated phase at late times dominated by dark energy. The significance of this transition  is robust to the choice of the underlying model.

\section{Conclusions}
We have shown how distance measurements from baryon acoustic oscillations (BAO) combined with distance moduli from Type-Ia supernovae (SN1a) can be used as a  cosmic distance ladder. This ladder can be calibrated at $z\sim 0$ using local determinations of the Hubble constant or at high redshift using the CMB determination of the sound horizon at radiation drag. The first approach is the classic (direct) cosmic distance ladder calibration while we refer to the second as an inverse cosmic distance ladder.
While the direct calibration is affected by a host of astrophysical  processes  it is  cosmological-model independent. The inverse ladder has much smaller calibration errors  if the early  ($z >~ 1000$) expansion history is standard,  but it is model-dependent.

In particular, we find that BAO and SN1a are quite complementary. SN1a luminosity distance data constrain very well the shape of the expansion history and they finely probe the redshift range  $0<z<1.3$  so that the  shape of the expansion history, $E(z)$, is very well constrained but  the overall normalization must be set externally for example by a direct determination of $H_0$. BAO on the other hand  cover sparsely the redshift range but  can be used to tie in the low redshift universe  to the high redshift one as the standard ruler is set at radiation drag ($z$ of ${\cal O} (1000)$).

The comparison between the two approaches is useful  for two purposes.
{\it i)} explore the origin of possible discrepancies between the cosmological constraints from the CMB and the ones derived from local measurements of the expansion rate. This approach is useful to disentangle information coming from the early Universe and  from the late one which are governed by  different physical processes. Comparing  early-time vs late time constraints has been and will continue to be an insightful way to probe new physics beyond the adopted cosmological model.
{\it ii)} to map directly the expansion history of the Universe.

We have presented the  reconstructed expansion histories derived by this combination of datasets for different cosmologies, and we find them to be very stable (both in shape and uncertainties), even when the curvature of the Universe or when the equation of state of dark energy are left as free parameters. 

By calibrating the BAO+SN1a cosmic distance ladder on the  sound horizon at radiation drag, $r_d$, we obtain  a robust determination of $H_0$ of 67.7$\pm$1.1 km s$^{-1}$ Mpc$^{-1}$.
A similar result is presented in \cite{BOSSBAO} by the BOSS collaboration, in which they study cosmological models with more degrees of freedom than here, i.e. in a general polynomial form of $H(z)^2$ that depends on $(1+z)^{\alpha}$, with $\alpha$=0,1,2,3. The results are completely consistent with those presented here. This is also true when comparing to \cite{Heavens2014}, whose assumptions are completely generic. Using slightly different BAO data they find $r_d$=142.3$\pm$6.1Mpc which is within 1$\sigma$ of our result for the O$\Lambda$CDM case. Overall, there is a broad agreement between the different analyses, implying that there seems to be no indication for deviations from $\Lambda$CDM, despite the different approaches and family of deviations considered.

On the other hand we can calibrate the same ladder on local measurements of $H_0$ obtaining  a constraint on $r_d$  which is independent on assumptions about early time physics and  early expansion history.  We find 137.0$\pm$5.0~Mpc for $\Lambda$CDM, and similar results for O$\Lambda$CDM, $w$CDM, and $ N_{\rm eff}\Lambda$CDM, as shown in Table~\ref{tab:rs}. This measurement is only weakly dependent on the assumed model for the late-time expansion history and on the assumed geometry.  The determination of $r_d$ is  consistent with the value measured by Planck of $147.5\pm 0.6$~Mpc for a standard cosmology with three neutrinos. Conversely, this measurement of $r_d$ places a limit on the number of relativistic species $N_{\rm eff}$ of 4.62$\pm$0.88. 
While with current CMB data there is still a degeneracy between $N_{\rm eff}$ and other cosmological parameters which propagates into a large uncertainty in $r_d$ for this model, this could be resolved by better measurements of the CMB damping tail and better peak localisation in the (E-mode) polarisation \citep[e.g.,][]{Houetal}.

With all of the above,  we find currently no compelling evidence to invoke non-standard cosmological models to explain the expansion history between redshifts $0<z<1.3$. The modest difference ($\sim 2 \sigma$) between the value of the Hubble constant measured directly and that inferred from the BAO+SN1a+$r_d$ ladder (in the context of $\Lambda$CDM, $w$CDM or O$\Lambda$CDM) and likewise the difference ($\sim 2 \sigma$) in the sound horizon measured by Planck and inferred from the BAO+SN1a+$H_0$ ladder may easily result from chance.  Indeed, substitution of WMAP data for Planck reduces this discrepancy further \citep{Bennett2014}. However, if the significance of this difference in future experiments rose above chance and beyond the reach of their systematic errors the approach illustrated here of comparing  direct and inverse distance ladders could provide evidence for new physics.

\section*{Acknowledgements}
AJC and LV are supported by supported by the European Research Council under the European Community's Seventh Framework Programme FP7-IDEAS-Phys.LSS 240117. LV and RJ acknowledge Mineco grant FPA2011-29678-C02-02. Based on observations obtained with Planck (http://www.esa.int/Planck), an ESA science mission with instruments and contributions directly funded by ESA Member States, NASA, and Canada. Funding for SDSS-III has been provided by the Alfred P. Sloan Foundation, the Participating Institutions, the National Science Foundation, and the U.S. Department of Energy Office of Science. The SDSS-III web site is http://www.sdss3.org/.

\bibliographystyle{mn2e}
\bibliography{ladder}

\end{document}